\documentclass[%
reprint,
 amsmath,amssymb,
 aps,
]{revtex4-2}

\usepackage{graphicx}
\usepackage{dcolumn}
\usepackage{bm}
\usepackage{hyperref}
\usepackage{multirow}

\begin{document}


\title{Using Convolutional Neural Networks to Accelerate 3D Coherent Synchrotron Radiation Computations }
\author{Christopher Leon}
\email{cleon@lanl.gov}

\author{Petr M. Anisimov}
\author{Nikolai Yampolsky}
\author{Alexander Scheinker}

\affiliation{Los Alamos National Laboratory, Los Alamos, NM 87545, USA}

\date{\today}

\begin{abstract}
Calculating the effects of Coherent Synchrotron Radiation (CSR) is one of the most computationally expensive tasks in accelerator physics. Here, we use convolutional neural networks (CNN's), along with a latent conditional diffusion (LCD) model, trained on physics-based simulations to speed up calculations. Specifically, we produce the 3D CSR wakefields generated by electron bunches in circular orbit in the steady-state condition. Two datasets are used for training and testing the models: wakefields generated by three-dimensional Gaussian electron distributions and wakefields from a sum of up to 25 three-dimensional Gaussian distributions. The CNN's are able to accurately produce the 3D wakefields $\sim$250-1000 times faster than the numerical calculations, while the LCD has a gain of a factor of $\sim$34. We also test the extrapolation and out-of-distribution generalization ability of the models. They generalize well on distributions with larger spreads than what they were trained on, but struggle with smaller spreads.

\end{abstract}
\maketitle

\section{Introduction}

Coherent Synchrotron Radiation (CSR) is a collective effect that can significantly distort an electron bunch undergoing circular motion. This occurs in the magnetic dipoles found in accelerator sections, such as the magnetic dipoles of storage rings and bunch compressors. Bunch lengths have continually shortened for applications like X-ray free electron lasers (XFEL's), which produce extremely bright X-ray pulses to probe matter. This has been tremendously useful for numerous fields, like the study of protein structure, imaging of viruses and matter at high energy density. This has led to the need to properly take into account the effects of CSR. However, calculations involving CSR can be extremely computationally expensive. Historically, 1D approximations have been widely used, but those are expected to break down eventually, specifically for large transverse-to-longitudnal ratios \cite{Derbenev:1995px}. The need for quick, 3D calculations for CSR has become important to the accelerator community. The evolution of computational hardware along development of new techniques opens up opportunities to seek alternative routes to the traditional methods. In this work, we get the wakefields used for CSR from machine learning (ML) surrogate models based on a convolutional neural network (CNN) architectures, as well as a latent conditional diffusion model. These were all trained on the physics-based software \texttt{PyCSR3D} \cite{mayes2021computational, christopher_mayes_2021_5496096}. The goal is to create ML models that are accurate while running much faster than the physics-based simulations.

\subsection{The Physics of CSR}

\begin{figure*}
\centering
\includegraphics[ width=0.8\textwidth ]{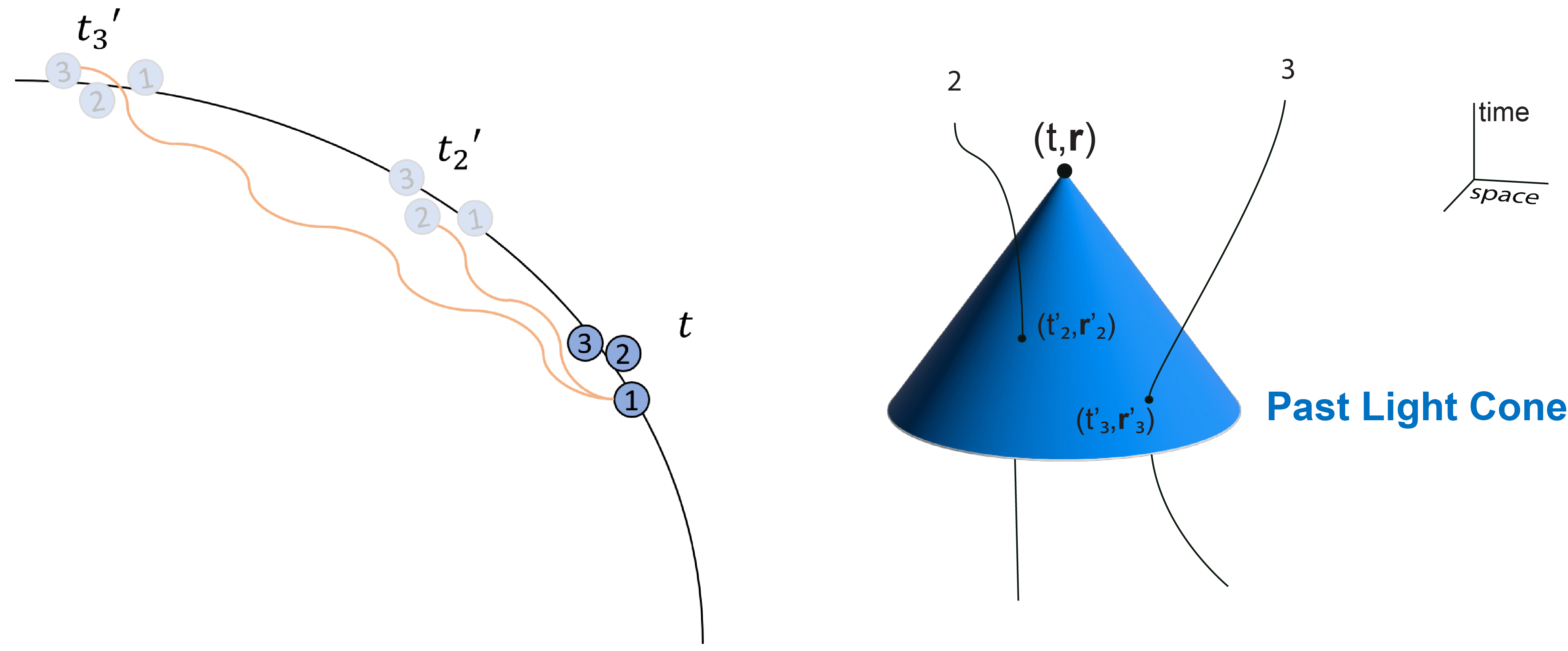}
\caption{ Three particles in a circular bend. \textit{Left}:  At time $t$, Particle 1 absorbs radiation emitted by Particles 2 and 3 at earlier points in time, $t_2'$ and $t_3'$, respectively. \textit{Right}: To get the total force a charged particle experiences at a particular point in space-time, $(t, \mathbf{r})$, one integrates the Lorentz force from the LW fields along the past light cone. This is called the wakefield, $\mathbf{W}(\mathbf{r}, t)$. }
 \label{CSR-cartoon}
\end{figure*}

To see the physical origins of the CSR phenomenon, first picture $N_e$ 
electrons in a bunch of length $\sigma$ undergoing circular motion in an external dipole magnetic field (see Fig. \ref{CSR-cartoon}). For the radiation emitted by the electrons, consider the part of the spectrum where the wavelength, $\lambda$, is  much larger than the bunch length: $\lambda \gg \sigma$. At $\lambda$ resolution, the bunch looks point-like, and thus the radiation emitted by the electrons at a given time is approximately in phase. This radiation adds up coherently and, given that in accelerators $N_e \sim 10^{10}$, the cumulative effect can be substantial. Even though the electrons are moving near the speed of light due to their small mass, they are travelling on a circular arc while the radiation travels linearly (see Fig. \ref{CSR-cartoon}). This means it's possible for there to be significant self-interaction for the bunch, which results in serious distortions of its phase space distribution. This generally has the effect of the tail and center losing energy, the head gaining, and an unwanted increase in emittance (volume in phase space).

To calculate the effect of CSR, one can start with the electromagnetic (EM) fields generated by a single relativistic point particle undergoing arbitrary motion. These are the well-known Liénard–Wiechert (LW) fields, \cite{lienard1898champ, wiechert1901elektrodynamische, jackson1999classical}:
\begin{equation}\label{lw-electric}
\mathbf{E} =  \left . -e \left( \underbrace{ \frac{\mathbf{n}-\boldsymbol{\beta }}{\gamma^2 (1-\mathbf{n}\cdot \boldsymbol{\beta})^3 \rho^2}} _\text{Coulomb term}
+ \underbrace{\frac{\mathbf{n} \times (\mathbf{n}-\boldsymbol{\beta }) \times \dot{\boldsymbol{\beta }}}{c(1-\mathbf{n}\cdot \boldsymbol{\beta})^3 \rho}}_\text{Synchrotron radiation} \right) \right |_{ret.},
\end{equation}

\begin{equation}\label{lw-magnetic}
\mathbf{B} = \mathbf{n}\times \mathbf{E},
\end{equation}
where Gaussian units are being used. Here, $\mathbf{E}$ is the electric field, $\mathbf{B}$ the magnetic field, $-e$ is the charge of the electron, $\mathbf{n}$ the normal vector from the source to the point of interest,  $\boldsymbol{\beta}$ the velocity normalized by the speed of light, $c$, $\gamma =\frac{1}{\sqrt{1-|\boldsymbol{\beta}|^2 }}$, and $\rho$ is the distance between the source and the point of interest. As seen in Eq. (\ref{lw-electric}), there are two terms: the Coulomb term, which gives the familiar Coulomb's law when the particle is at rest, and the synchrotron radiation term.

The $ret.$  in Eq. (\ref{lw-electric}) signifies that the expression must be evaluated at the retarded time, 
\begin{equation}\label{retarded-time}
 t' = t - \rho/c,   
\end{equation}
to ensure causality. That is, the EM fields at a point in time depend on the motion of the electron at the time of emission (see Fig. \ref{CSR-cartoon}).

From the LW fields, the wakefield, $\textbf{W} (\textbf{r}, t)$, the force a point particle experiences at a point in spacetime from the EM fields generated from other particles at earlier times, can be obtained. Getting  the Lorentz force, $\textbf{F} (\textbf{r}, t) = -e \left( \mathbf{E}(\textbf{r}, t) + \mathbf{v} \times \mathbf{B}(\textbf{r}, t)\right)$, along with Eq.'s (\ref{lw-electric}) and (\ref{lw-magnetic}), then integrate along the past light cone:
\begin{equation}
\textbf{W} (\textbf{r}, t) = \int_\triangle d^3 \mathbf{r'} \lambda(\mathbf{r'}, t') \mathbf{F}(\mathbf{r'} - \mathbf{r}, t') \quad 
\end{equation} 
where $\lambda(\mathbf{r'} , t')$ is the number density and $\triangle$ signifies the past light cone. We assume the radiation reaction force on the emitting electron is negligible, and it will not be considered here (see Baez \cite{Baez:2016snx} for an overview of the thorny theoretical issues and review of the works of Abraham, Dirac, and others on the topic). Doing the calculation naively, one would go into the past light cone for each particle, sum up all LW fields produced, and use the Lorentz force to get the wakefield (see Fig. \ref{CSR-cartoon}). This would be extremely inefficient,  as it would require storing the position, velocity and acceleration of each of the $N_e$ particles for each time step, as well doing computationally expensive $\mathcal{O}(N_e^2)$ calculations for each time step. Much of the previous work on CSR has focused on creating and implementing approximate models to make the challenging task of calculating its effects on electron beams computationally tractable. 

\subsection{Past Work}\label{previous-work}

Here, we provide a short review of past work done on CSR. For a comprehensive review of the CSR literature and the different techniques used, see Mayes \cite{mayes2021computational}. 

\begin{figure}
\centering
\includegraphics[ width=0.95\columnwidth ]{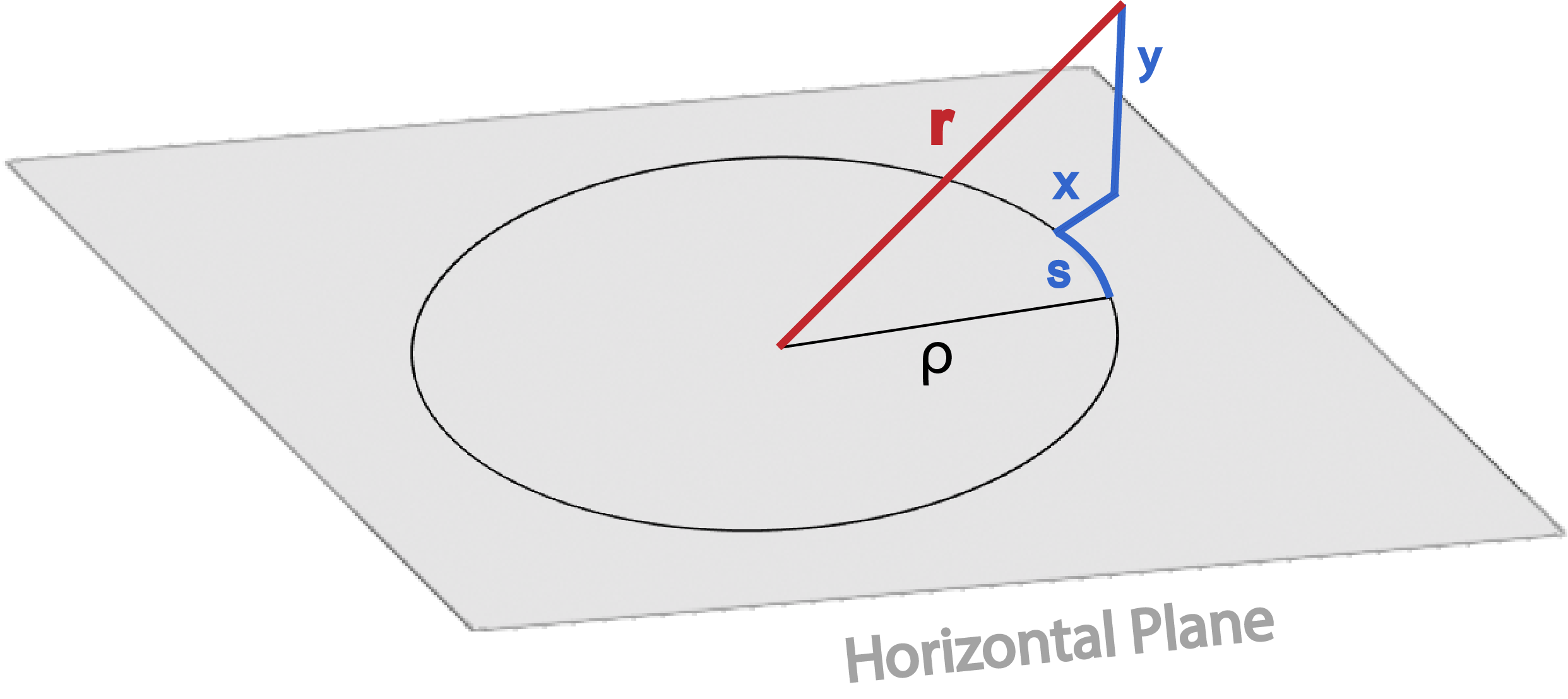}
\caption{Illustration of the Frenet-Serret coordinates for a position, $\boldsymbol{r}$. The reference path here is uniform circular motion with a circle of radius $\rho$ in the horizontal plane.}
 \label{Frenet-Serret}
\end{figure}

Raubenheimer and Carlsten calculated the emittance growth in circular accelerators  \cite{Carlsten:1995uj}. A small angle, ultra-relativistic, 1D model that treats the bunch as a line charge distribution, $\lambda_{1D}(s)$, and included transient effects was created by Saldin et al. \cite{Saldin:1996gs}. It was later expanded in \cite{Stupakov:2002xs}. Like many examples of using only using effective degrees of the freedom, complications arose from infinities. The issue was dealt with by `renormalizing'  the Coulomb term in Eq. (\ref{lw-electric}); that is, by subtracting the field generated during uniform motion. In the 1D steady-state case, the longitudinal wakefield in Frenet-Seret coordinates (see Fig. (\ref{Frenet-Serret})) is then, 
\begin{align}\label{1D-theory-Ws}
W_s(s) = - \frac{e}{3^{1/3}\rho^{5/3}} \int_{-\infty}^s ds' \frac{\partial \lambda_{1D} (s')}{\partial s'} \frac{1}{(s-s')^{1/3}}  \\ \text{(1D theory)}, \nonumber
\end{align}
with the transient case being a slight modification of Eq. (\ref{1D-theory-Ws}). The commonly used software \texttt{ELEGANT}  (ELEctron Generation ANd Tracking) \cite{borland2000elegant} uses this method to calculate the effects of CSR \cite{Borland:2001xua}. A bend is broken up to a series of segments and the effect of Eq. (\ref{1D-theory-Ws}) is calculated for each, with the approximation of the distribution being used for the calculation is the distribution at that point. This holds in the ultra-relativistic limit, since the radiation is concentrated in the forward direction with an angular spread of $\sim 1/\gamma$.


In 2D and 3D, the infinity issue and the need to renormalize does not come up. 
A two-dimensional model of CSR making use of Gaussians was created,  \texttt{CSR Track} \cite{Dohlus:2004ax, dohlus2007csrtrack}, which approximates a distribution as a sum of Gaussian sub-bunches. A 2D model of CSR was also created by Huang, Kwan, and Carlsten \cite{Huang:2013gr}. LW3D is a result of Ryne et al. using the LW fields and a particle based approach \cite{Mitchell:2012zz, Ryne:2018gfu}. It's been found that the effects of the Coulomb term can be significant during transience. As the bunch enters the bend, a pancaked EM field, originating from the free relativistic electron bunch in the drift section, continues `moving' into the bend  and can interact with the bunch inside  \cite{novokhatski2012coherent, Brynes:2018khp}. 

Recently, in a series of papers \cite{Cai:2017qui, Cai:2020kvm, Cai:2021xxm}, Cai created models for CSR in a bend utilizing the Hamiltonian formalism in Court-Synder theory \cite{Courant:1958wbj}. This was first done for a 2D model for steady state \cite{Cai:2017qui}, then with Ding for the 3D case in steady state \cite{Cai:2020kvm} and finally in the 2D case with transient effects \cite{Cai:2021xxm}. Applying the formalism to the first bunch compressor at the Linac Coherent Light Source at SLAC, Cai and Ding's method accurately captured the change in the emittance. Furthermore, it predicted a change in the vertical emittance, whereas \texttt{ELEGANT} had constant vertical emittance during a bend \cite{Cai:2020kvm}. 

For the problem of CSR, Frenet-Serret co-ordinates for a particle undergoing uniform circular motion in the horizontal plane are typically used. The longitudinal direction here is the circle arc length, $s$, the horizontal $x$ corresponds to the radial distance from the reference circle, and the vertical is denoted $y$ (see Fig. \ref{Frenet-Serret}). In the Snyder-Courant approach \cite{Courant:1958wbj, Stupakov:2018ove}, $s$ becomes the `time' axis, and derivatives with respect $s$ will be denoted with a prime (e.g. $x'$). In this approach, the equations of motion for a single particle are \cite{Cai:2020kvm}:

\begin{equation}\label{delta-equation}
\delta' = \frac{r_e N_b}{\gamma} W_s,
\end{equation}

\begin{equation}\label{x-equation}
x'' = \frac{r_e N_b}{\gamma} W_x,
\end{equation}

\begin{equation}\label{y-equation}
y'' = \frac{r_e N_b}{\gamma} W_y,
\end{equation}
where $r_e$ is the classical electron radius ($r_e =\frac{e^2}{m_ec^2}$) and $\delta =(p -p_0)/p_0$ where $p_0$ is the reference momentum ($p_0 = \frac{ e \rho B_0}{c}$). From Eq.'s (\ref{delta-equation}) - (\ref{y-equation}) one can see that the wakefields have units of $1/m^2$. 

In Cai's approach, shielding, where larger wavelengths get suppressed from the surrounding metallic beam pipe, was considered negligible in all cases. The paraxial approximation is also used in places to make the problem more tractable. To solve for the retarded time condition in Eq. (\ref{retarded-time}), a fourth order approximation was used, which is computationally faster than a root finding algorithm since the fourth order solution is analytic, and it was shown to be quite accurate as well \cite{Cai:2017qui, Cai:2020kvm}. 

In practice,  a combination of analytic models and numerical software can be used to minimize the effects of CSR. Xu et al. demonstrated the reduction of horizontal momentum shift in a bunch compressor for the proposed MaRIE FEL using a double chicane, where the momentum shifts produced by each chicane largely cancel each other out \cite{Xu:2022epa, xu2023x}. A coarse optimization was first done using a fast, simple analytical model, followed by refinement with a slower, more accurate simulation. 
The end result preserved beam emittance well. One can use the same approach for general CSR applications, but replace the analytic model with an ML surrogate model.

\subsection{Convolutional Neural Networks}

CNNs, originally designed for image processing, use locality in the data to build higher-level features \cite{lecun1989backpropagation}. CNNs also possess the inductive bias of translation equivariance, which many physical systems possess; colloquially, `the laws of physics are the same everywhere'. Locality and translation equivariance make CNNs well-suited ML surrogate models for various physical systems.

While using ML surrogate models can have a significant upfront time cost for training the models on simulation data, once the training is done, evaluations can be performed extremely quickly. This speed is due to evaluations leveraging matrix multiplication and parallelized operations optimized by modern GPUs. CNNs have been used elsewhere to replace computationally intensive physics-based simulations with ML surrogates. For example, in \cite{Scheinker:2023onb} 3D CNN’s for electrodynamics with hard physics constraints were first developed by creating neural operators that map charge and current densities directly to their associated scalar and vector potentials from which the electric and magnetic fields are generated according to Maxwell’s equations. In \cite{Leon:2024wsz}, a Fourier-Helmholtz-Maxwell Neural Operator (FoHM-NO) approach was developed which incorporates almost all of Maxwell’s equations as hard physics constraints and can accurately simulate the electromagnetic fields generated by charged particle beams. Both of these approaches were shown to reduce evaluation time from approximately two days when using the General Particle Tracker simulations to just a few seconds for AI model-based inference.

Previous work has investigated the use of ML surrogate models to speed up CSR calculations. Mayes and Edelen used a dense neural network in the 1D case to speed up CSR calculations \cite{Edelen:2022vij}. CNN's have been utilized before in the 2D transient case \cite{Robles:2023tus}. Gonzalez-Aguilera et al. used generative methods to reconstruct projections of phase space affected by CSR \cite{Gonzalez-Aguilera:2024ewh}. To our knowledge, this work is the first in-depth study of employing CNN surrogate ML models to calculate the full 3D CSR wakefields.

\section{Training and Test Data}

\begin{table}[]
\centering
 \label{datasets-table}
\begin{tabular}{lll}
\hline
                                                                   & \begin{tabular}[c]{@{}l@{}}\textbf{Gaussian} \\ \textbf{Dataset}\end{tabular}                                                                  & \begin{tabular}[c]{@{}l@{}}\textbf{Multi-Gaussian} \\ \textbf{Dataset}\end{tabular}                                                                   \\ \hline
\begin{tabular}[c]{@{}l@{}}Physical \\ Range\end{tabular}          & \begin{tabular}[c]{@{}l@{}}$x_i \in (-48, 48)~\mu m$ \\ $i = x, y, s$\end{tabular}                                              & \begin{tabular}[c]{@{}l@{}}Same as \\ Gaussian Dataset\end{tabular}                                                          \\ \hline
\begin{tabular}[c]{@{}l@{}}Distribution \\ Attributes\end{tabular} & \begin{tabular}[c]{@{}l@{}}$\sigma_i \in (2, 12)~\mu m$ \\ $i = x, y, s$ \\ Centered at $\textbf{0}$\end{tabular} & \begin{tabular}[c]{@{}l@{}}Sum of 2 to 25 \\ Gaussians \\ $\mu_i \in (-12, 12) ~\mu m$ \\ $\sigma_i \in (2, 12) ~\mu m$ \end{tabular}
\end{tabular}
\caption{Physical characteristics of the two datasets. Distributions are created by being uniformly sampled in the ranges here.} \label{data-details}
\end{table}

We generated the training data for the surrogate ML model using the software package \texttt{PyCSR3D} \cite{christopher_mayes_2021_5496096, mayes2021computational}, which produces the 3D wakefield of a given distribution undergoing circular motion in the steady state case. The physics of the software is based on the model by Cai and Ding \cite{Cai:2020kvm} mentioned in Section \ref{previous-work} and includes the Coulomb term. The choice of \texttt{PyCSR3D} was because in a study it outperformed several other mesh-based methods \cite{mayes2021computational}, specifically by most accurately reproducing the entire 3D wakefield, where ground truth was taken to be the results from the far more computationally expensive, particle-based LW3D \cite{Ryne:2018gfu}. Computationally, \texttt{PyCSR3D} uses the Integrated Green Function approach to get the transverse wakefields. Convolutional integrals are evaluated numerically by performing Fast Fourier Transforms, utilizing the convolution theorem, and then performing an inverse Fast Fourier Transform.

Two datasets were created, the Gaussian Dataset and the Multi-Gaussian Dataset. The first dataset is used to evaluate the ML-based approach relative to how simple Gaussian distribution-based models approximate real world conditions. The second assesses the models against more complex distributions that may also show up. Both datasets consisted of bunches in a bend with $\rho = 1$ m and $\gamma = 500$. The data is on a uniform spaced grid in Frenet-Serret coordinates, the physical characteristics of which can be seen in Table \ref{data-details}.

\subsection{Gaussian Dataset}
The Gaussian Dataset consisted of 500 Gaussian distributions centered at $\mathbf{0}$, their standard deviations and their corresponding wakefields.  Field configurations are represented as an array of $(128, 128, 256, n_F)$, where $n_F =1$ ($n_F =3$) for a scalar (vector) field. The third index represents the $s$ component. It has twice the resolution of the transverse components, the first and second indices, as the longitudinal coordinate is usually the most important. 

Starting with Gaussians distributions is useful as there are a number of analytic results from approximate models to compare results to, thus they serve as good sanity checks on the data and models. Real beams can also sometimes approximate Gaussians and are often modeled as such. 

\subsection{Multi-Gaussian Dataset}

Each distribution in the Multi-Gaussian Dataset is a sum of Gaussians, with the exact number, $N_G$, being chosen randomly and uniformly between 2 and 25 for each distribution. The total number of particles is the same as in the Gaussian Dataset, and each individual Gaussian has a $1/N_G$ fraction of the particles. There are a total of 500 examples of multi-Gaussian distribution/wake pairs.

The means and standard deviations of each individual Gaussian were chosen randomly and uniformly to be inside some bounds (see Table \ref{data-details} for exact details). Approximating a distribution as a sum of Gaussians is commonly used in Kernel Density Estimation. 
More physically, something like this is seen in micro-bunching. 

For both the Gaussian and Multi-Gaussian sets, the scale of $W_s$ was $\sim 10^6 \frac{1}{m^2}$, the scale of $W_x$ about an order of magnitude smaller  and that of  $W_y$ was about two orders of magnitude smaller. To avoid $W_s$ dominating completely when training, both the Gaussian and Multi-Gaussian Datasets had each component normalized separately, by the means and standard deviations of the components of the $W_i$ in the Gaussian Dataset. 

\section{Models}

There is significant flexibility in choosing the ML architecture and generality of the models. Thus, to explore the performances across different cases, we consider three machine learning architectures. The $\sigma$-NN is the simplest of the three with only a decoder CNN architecture, while also having the most restrictive use case of just Gaussian distributions. The $\lambda$-NN has a modified U-Net architecture, which is computationally slower than $\sigma$-NN, but lends itself to more general use. Finally, the LCD uses state-of-the latent diffusion and has a similar general use case as the $\lambda$-NN.  The following subsections describe the models in more details.

\subsection{Sigma Model}

\begin{figure*}
\centering
\includegraphics[ width=0.85\textwidth ]{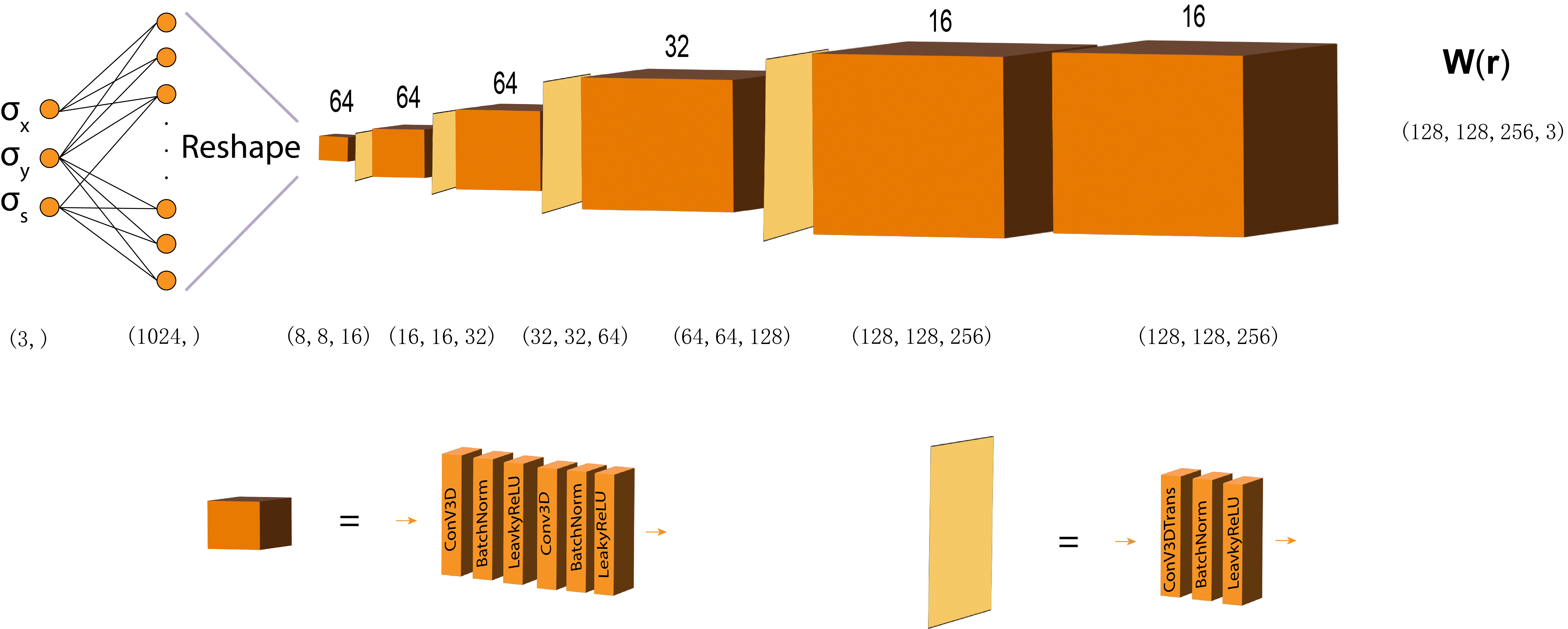}
\caption{\textbf{$\sigma$-NN architecture} -  A dense network is used to create a 1024 dimensional vector, which is reshaped to a (8,8,16) array. The output is processed through several convolutional and convolutional transpose block layers. Numbers over blocks represent the number of filters.}
 \label{Sigma-NN-architecture}
\end{figure*}

The $\sigma$-NN model produces the wakefield of Gaussian density distributions given the spreads in $x$, $y$, and $s$:
\begin{equation} 
\sigma\text{-NN}: \ \
(\sigma_x, \sigma_y, \sigma_s) \rightarrow \mathbf{W}(\mathbf{r}).
\end{equation}
In terms of arrays of the dataset, $(3,) \rightarrow (128, 128, 256, 3)$.    

The $\sigma$-NN has a decoder CNN architecture (see Fig. \ref{Sigma-NN-architecture}). 
The model has the built-in bias that the three $\sigma_i$ parameters describe the entire wakefield. This limits its applicability, but it is expected to perform better when the bunch is Gaussian. It also follows the tradition of software like \texttt{CSRTrack} \cite{Dohlus:2004ax} that calculates the wakefield of a general distribution using approximation of Gaussian distributions. $\sigma$-NN  was only trained and tested on the Gaussian Dataset.

\subsection{Lambda Model}

\begin{figure*}
\centering
\includegraphics[ width=0.97\textwidth ]{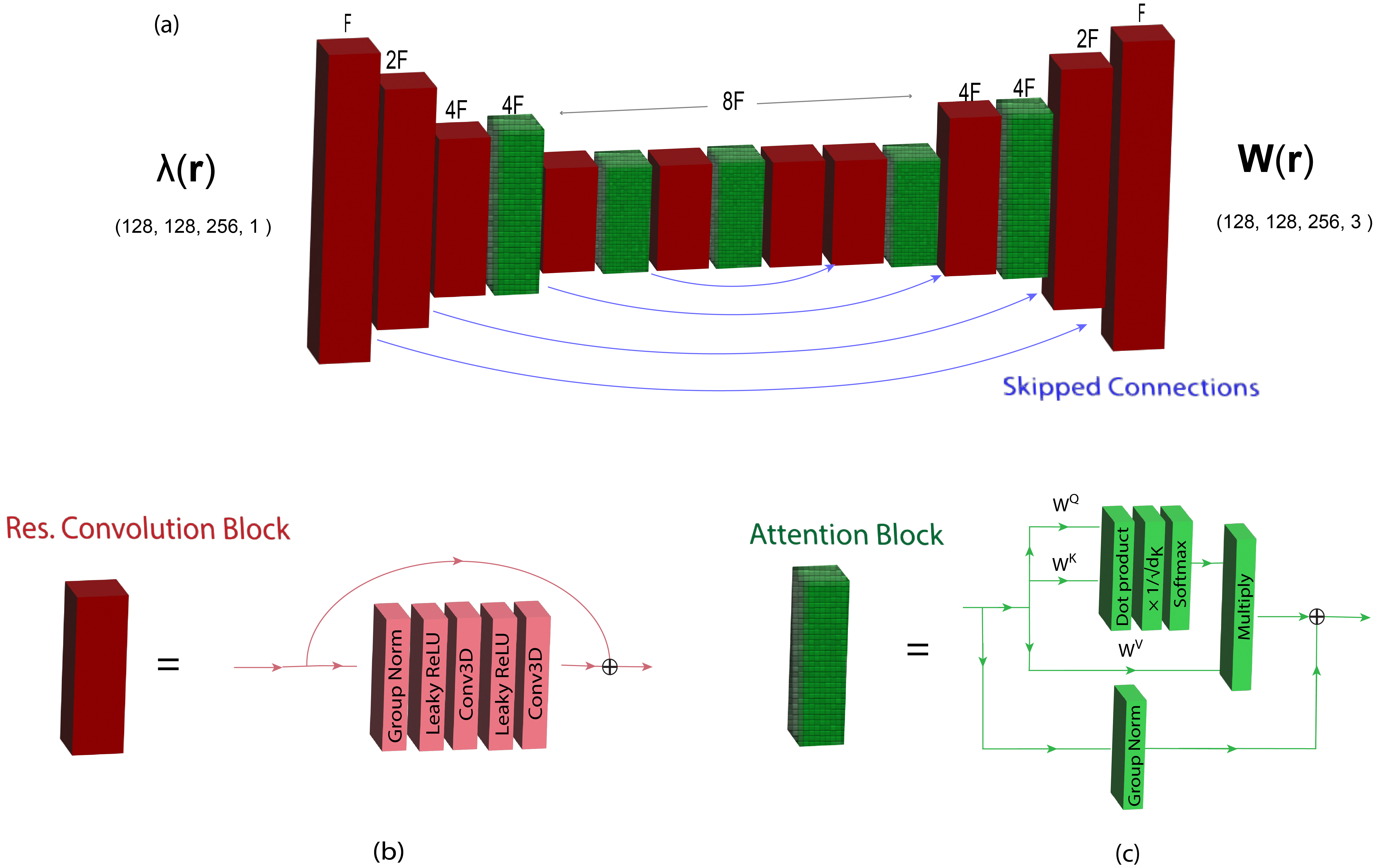}
\caption{ \textbf{$\lambda$-NN architecture}   (a) A U-Net architecture \cite{ronneberger2015u} constructed out of  residual convolution and attention blocks. A down/up sampling layer is implied for every time the block dimensions change.  $F$ is a parameter here, which was taken as $F=4$. (b)  These blocks are of a deep residual structure \cite{He:2015wrn}. The input gets sent to through two paths: one part goes through several layers, while on the other it gets operated on by the number of $1\times1\times1$ convolutions needed to make the dimensions the same for addition. The two outputs are then summed. $F$ here is the number of filters. (c) The Attention Blocks use the attention mechanism \cite{Vaswani:2017lxt} on the feature maps. For a single input, the query, key and value vectors, of dimension $F$, are obtained by multiplication by the matrices $W^Q, W^K$ and $W^V$, respectively. A weighted average of the value vector is taken, with the weights acquired through the use of softmax on the query and key vectors. This gets added to the group normed input. In practice, several inputs are fed simultaneously in order to train more efficiently. }
 \label{Lambda-NN-architecture}
\end{figure*}

\begin{figure*}
\centering
\includegraphics[ width=0.9\textwidth ]{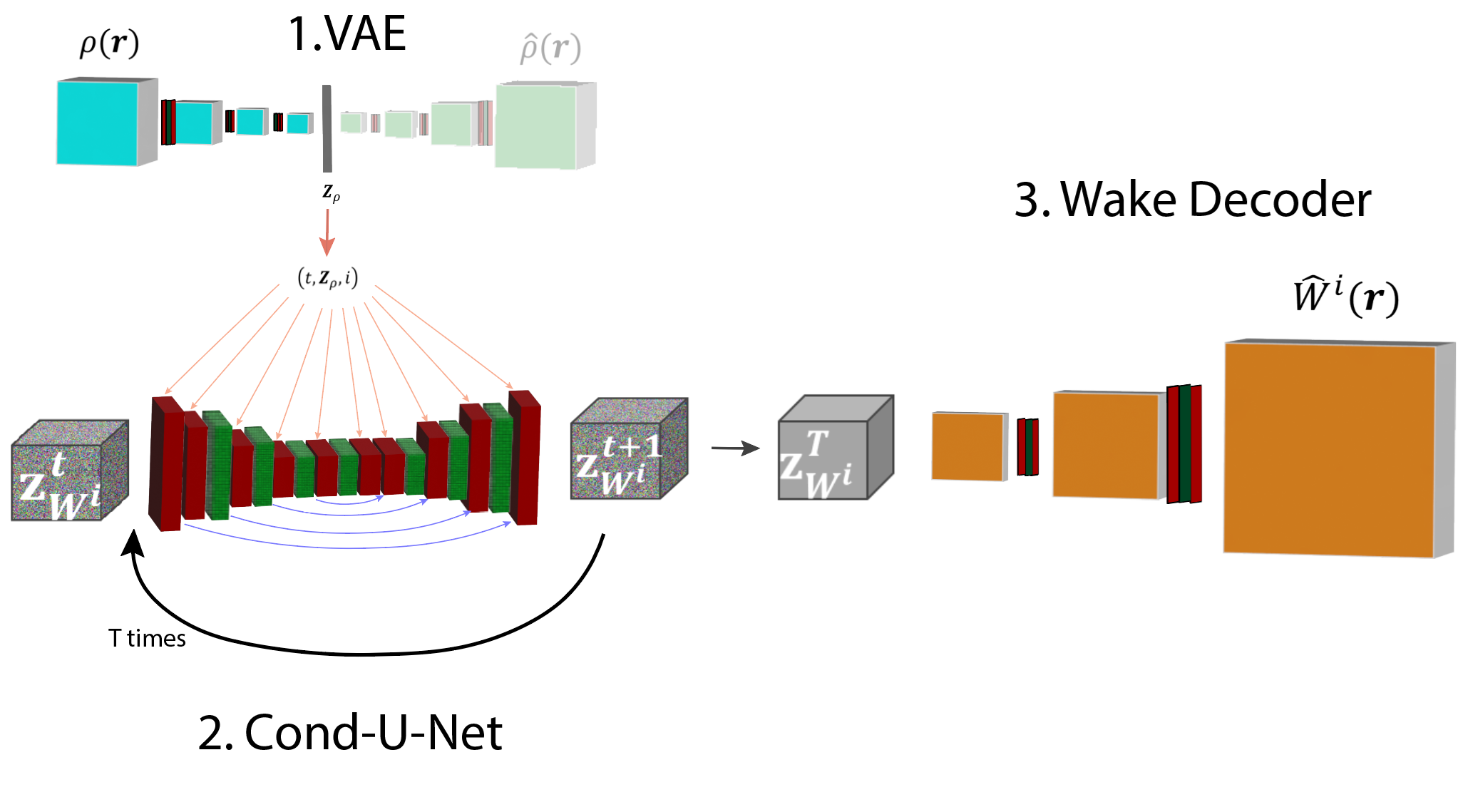}
\caption{The Conditional Latent Diffusion Model }
\label{Diffusion-architecture}
\end{figure*}

The $\lambda$-NN model is more general and produces the wakefield given an arbitrary 3D distribution, $\lambda(\mathbf{r})$,
\begin{equation}
\lambda \text{-NN}: \ \ \lambda(\mathbf{r}) \rightarrow \mathbf{W}(\mathbf{r}).
\end{equation}
In terms of arrays, $(128, 128, 256, 1) \rightarrow (128, 128, 256, 3)$. 

A U-Net architecture \cite{ronneberger2015u} was chosen for the $\lambda$-NN model (see Fig. \ref{Lambda-NN-architecture}). This builds on a CNN encoder-decoder architecture with the addition of skip connections, which concatenate layers in the encoder to same sized layers in the decoder. `Blurriness' in CNN encoder-decoder architectures is a well-known phenomenon and likely stems from the fact that much of the finer detailed information in the original input gets lost during the compression into the latent space. The skipped connections in the U-Net fix this by having the decoder also pull information from the encoder in a multiscale fashion (see Fig. \ref{Lambda-NN-architecture}). This gives the decoder more information, while still forcing the network to find important, salient features during the compression down to the latent space. In addition, the skipped connections also offer a benefit during the training process, as they offer a `gradient superhighway' that allows for better optimization during gradient descent, which is particularly beneficial for parameters that would otherwise be far removed from the output.

We follow Ref. \cite{ho2020denoising} using a U-Net with Res-Net block modules, which is based on PixelCNN++ \cite{salimans2017pixelcnn++} with group normalization \cite{wu2018group}. This modifies the U-Net with the addition of five attention layers— two in the encoder, two within the decoder, and one in the latent space. These attention layers utilize the self-attention mechanism, a fundamental aspect of transformers in large language models \cite{Vaswani:2017lxt}. These layers help the model learn global correlations using higher-level features.

We first trained $\lambda$-NN on the Gaussian Dataset, which we'll refer to as $\lambda$-NN-Gauss. It was taken as a pre-trained model and then trained on the Multi-Gaussian Dataset. This saved time over training from scratch and allowed us to explore how well learning transferred from going from one data set to another. The final resulting model we label $\lambda$-NN-Muli. 

\subsection{ Latent Conditional Diffusion Model}

Diffusion models are frequently used in modern image  generation \cite{ho2020denoising}, and are employed in technologies such as DALL-E and Stable Diffusion \cite{rombach2022high}. They have also been used in accelerator physics as generative virtual beam diagnostics to generate the phase space distribution of particle beams based on limited data \cite{Scheinker:2024gaj, Scheinker:2024rfl}, and have been found to have connection to the stochastic quantization of quantum lattice field theories \cite{Wang:2023exq}. Inspired by non-equilibrium thermodynamics \cite{Sohl-Dickstein:2015dhe}, these models draw analogies to fluid systems starting from low entropy and evolving to pure randomness. The data is corrupted by the successive addition of noise, until the image is effectively just pure Gaussian noise. The machine learning algorithm learns to denoise images step-by-step so that, starting from pure randomness, an image can be generated.

Specifically, in the forward process of diffusion, we start with an “image” of $\boldsymbol{Z}^0$ then have a Markov chain sequence defined by:
\begin{equation}
\boldsymbol{Z}^t = \sqrt{1-\beta_t}\boldsymbol{Z}^{t-1} + \sqrt{\beta_t} \boldsymbol{\epsilon}_t, \ \ \boldsymbol{\epsilon}_t \sim \mathcal{N}(\boldsymbol{0}, I),
\end{equation}
where $\{ \beta_t \}_{t=1}^T$ is a chosen `noise schedule' and  $0 < \beta_t < 1$. Noise comes to dominate in the sense that $\lim\limits_{t \to \infty} \boldsymbol{Z}^t \sim \mathcal{N}(\boldsymbol{0}, I)$. Given the analytical properties of multidimensional normal distributions, this makes the diffusion process mathematically tractable and computationally efficient. Subsequently, the network is trained with the backward diffusion process, learning to denoise by predicting $\boldsymbol{Z}^{t-1}$ given $\boldsymbol{Z}^{t}$. After training, starting from Gaussian noise, $\boldsymbol{Z}^0 \sim \mathcal{N}(\boldsymbol{0}, I)$, the model iteratively denoises to produce an image. This approach transforms the problem of sampling from a difficult, complex target distribution, such as $128\times128$ pixel images of cats, to a simpler one of sampling from a known prior distribution (Gaussian noise).

Latent diffusion \cite{rombach2022high} was specifically developed for cases when the data is high dimensional (e.g. high resolution images), and it becomes more computationally efficient to operate in a lower dimensional latent space representation. Conditioning is done by specifying aspects of the `image'. For example, using the text 'cat' to generate an image of a cat. Here, the $\rho(\boldsymbol{r})$ field is compressed down to a vector $\boldsymbol{Z}_{\rho}$, using a variational autoencoder (VAE) for conditioning. Since the field configuration for a single component of $W^i (\boldsymbol{r})$  are for us represented by large (128, 128, 256, 1) arrays, we will also work in the latent space. The structure and regularity in the data means it can be described by far fewer than $128\times128\times256 \sim 4 \times 10^6$ parameters. This is accomplished with another convolutional autoencoder, which compresses the wakefield down to an (8, 8, 16, 4) array, a compression of $\sim 0.1$ \%. Although in the Gaussian case the data can be compressed to the three $\sigma_i$ parameters, we keep the latent space larger in order to better represent a more general case. We condition also based on the ``time'' of the diffusion, latent density distribution vector $\boldsymbol{Z}_{\rho}$, and the component we wish to focus on, $i$, which is represented as a one hot encoding vector. This way, all three $W^i$ components can be calculated in parallel, but also only 1 or 2 components can be obtained if one wishes to use a 1D or 2D approximation.

In summary, the procedure of the LCD     model is processed through 3 submodels (see Fig. \ref{Diffusion-architecture}): 
\begin{enumerate}
    \item \textit{VAE}: compress $\rho(\boldsymbol{r})$ with the encoder of a VAE:
    \begin{equation}
        \rho(\boldsymbol{r})  \rightarrow \boldsymbol{Z}_{\rho} \nonumber
    \end{equation}

    \item \textit{Conditional U-Net with Attention Layers}: Starting with pure noise in the latent space of the wakefields, iteratively denoise using conditioning on time, $\boldsymbol{Z}_{\rho}$, and component $i$ :
    \begin{align}
    &\boldsymbol{Z}_{W^i}^0 \sim \mathcal{N}(\boldsymbol{0}, I) \nonumber \\
        &\text{for} \ \ t \in \{0, 1,...T-1\}: \nonumber\\
            & \quad \boldsymbol{Z}_{W^i}^{t}|(t, \boldsymbol{Z}_{\rho} , i) \rightarrow \boldsymbol{Z}_{W^i}^{t+1}
        \nonumber
    \end{align}

    \item \textit{Wake Decoder}: A previously trained decoder is used to get go from the wakefield latent space to produce an estimate of the wakefield:
    \begin{equation}
    \hat{\boldsymbol{Z}}_{W^i}^T \rightarrow \hat{\boldsymbol{W}}_i (\boldsymbol{r}) \nonumber
    \end{equation}
\end{enumerate}

For the conditional latent diffusion architecture, we also follow Ref. \cite{ho2020denoising} for the denoising model using a U-Net with Res-Net block modules. This model incorporates seven attention layers—three within the encoder, three within the decoder, and one in the latent space. For time embedding, we use the transformer sinusoidal position embedding \cite{Vaswani:2017lxt}. The noise schedule was $\beta_1 = 10^{-4}$,  $\beta_T = 0.5$, and $T = 12$.

\subsection{Architecture Hyperparameters and Training}
Many of the hyperparameters in the architectures and training were chosen to be powers of 2. This is recommended by Nvidia \cite{Nvidia-powers-of-2} for deep learning in order to most efficiently utilize its GPU's, which rely on dividing calculation operations into many parallelized processes. Filter sizes of $3\times3\times3$ are used in all the convolutional layers, as small filter sizes with more layers can provide the same coverage as larger filter sizes with fewer layers, but with fewer parameters. To smooth the predicted fields, we added a final averaging layer with a $3\times 3\times 3$ window and then trained for a few more epochs for $\sigma$-NN and $\lambda$-NN-Gauss. We found this significantly improved results.

For the $\sigma$ and $\lambda$ models, leaky ReLU's \cite{maas2013rectifier} with $\alpha = 0.1$ were used for the activation function, as they have the same advantages as the more commonly used ReLU, but suffer less from the vanishing gradient problem. The LCD model used the Swish activation function \cite{Ramachandran:2017jge}.

For training of all the models, the Adam optimizer \cite{kingma2014adam} was used, and the mean squared error (MSE) was chosen for the loss function. Both datasets had a 85/15 \% training/test split. The learning rate was decreased during the training process as to `turn down the noise' while optimizing. That is, there is high noise in the beginning to encourage greater exploration of parameter space and avoid getting stuck in local stationary points, while the noise is decreased towards the end as the best local set of parameters is honed in on. There are parallels here to simulated annealing, where the temperature acts as the source of the noise and is decreased during optimization to find the global minimum \cite{smith2017don}. TensorFlow \cite{abadi2015tensorflow} was used for the creation of the neural networks and their training. All computations were done on a workstation with an Intel Xenon Platinum 8268 processor at 2.90GHz and two NVIDIA RTX A6000 GPU's with 48 GB of RAM.

\section{Results}

\begin{figure}
\centering
\includegraphics[ width=0.98\columnwidth ]{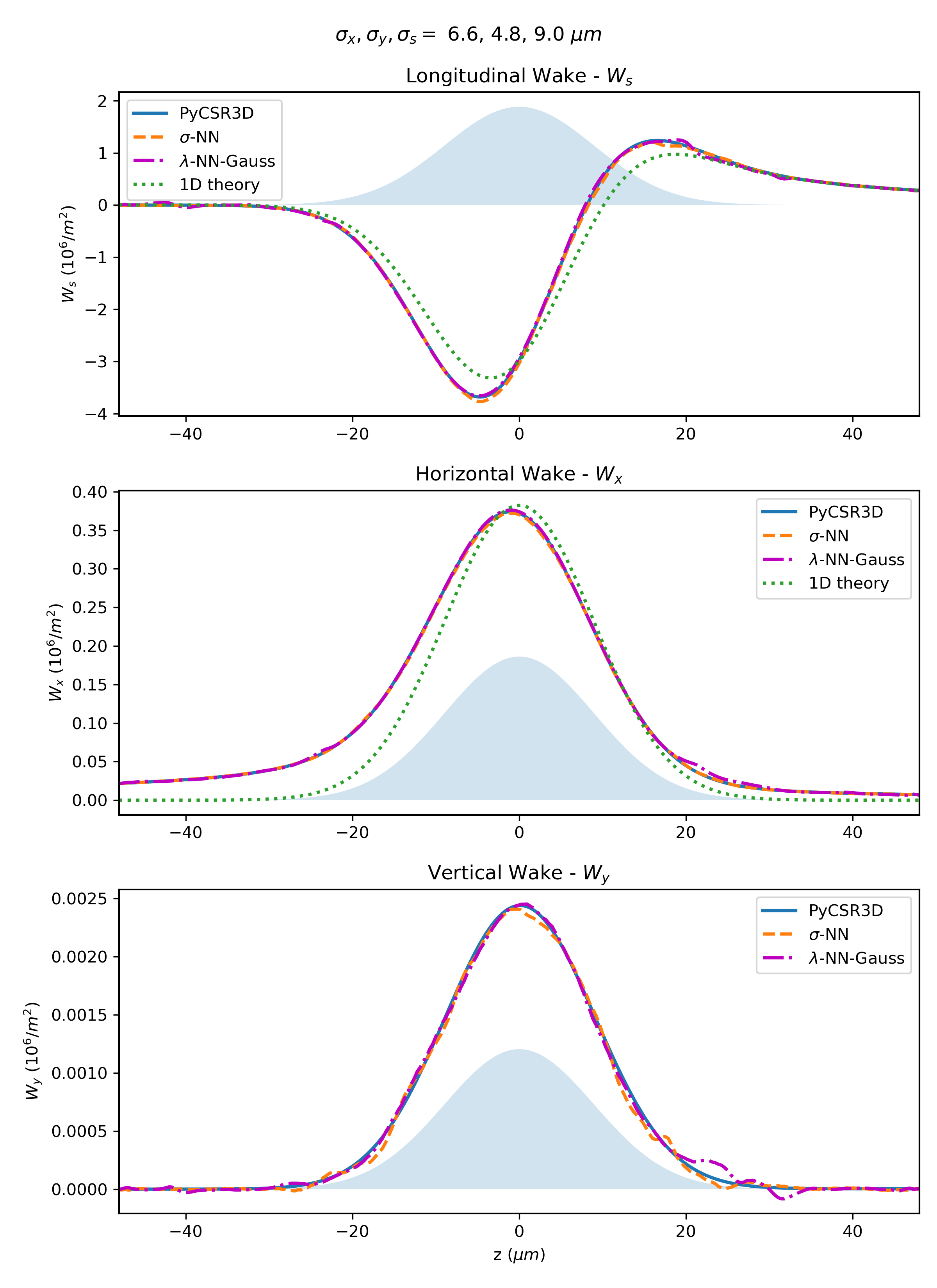}
\caption{\textbf{Gaussian Dataset}: The performance of $\sigma$-NN and  $\lambda$-NN-Gauss on a distribution in the Test Set. This is along the z axis ($x = y = 0 ~ \mu m$). The electron distribution is shown in the shaded blue area. The 1D theory for $W_s$ is given by Eq. (\ref{1D-theory-Ws}), while for $W_x$ it uses Eq.'s (\ref{Wx-1D-theory}) and (\ref{Lambda-used}).}
 \label{Sigma-Lambda-NN-1D}
\end{figure}

\begin{figure*}
\centering
\includegraphics[ width=0.95\textwidth ]{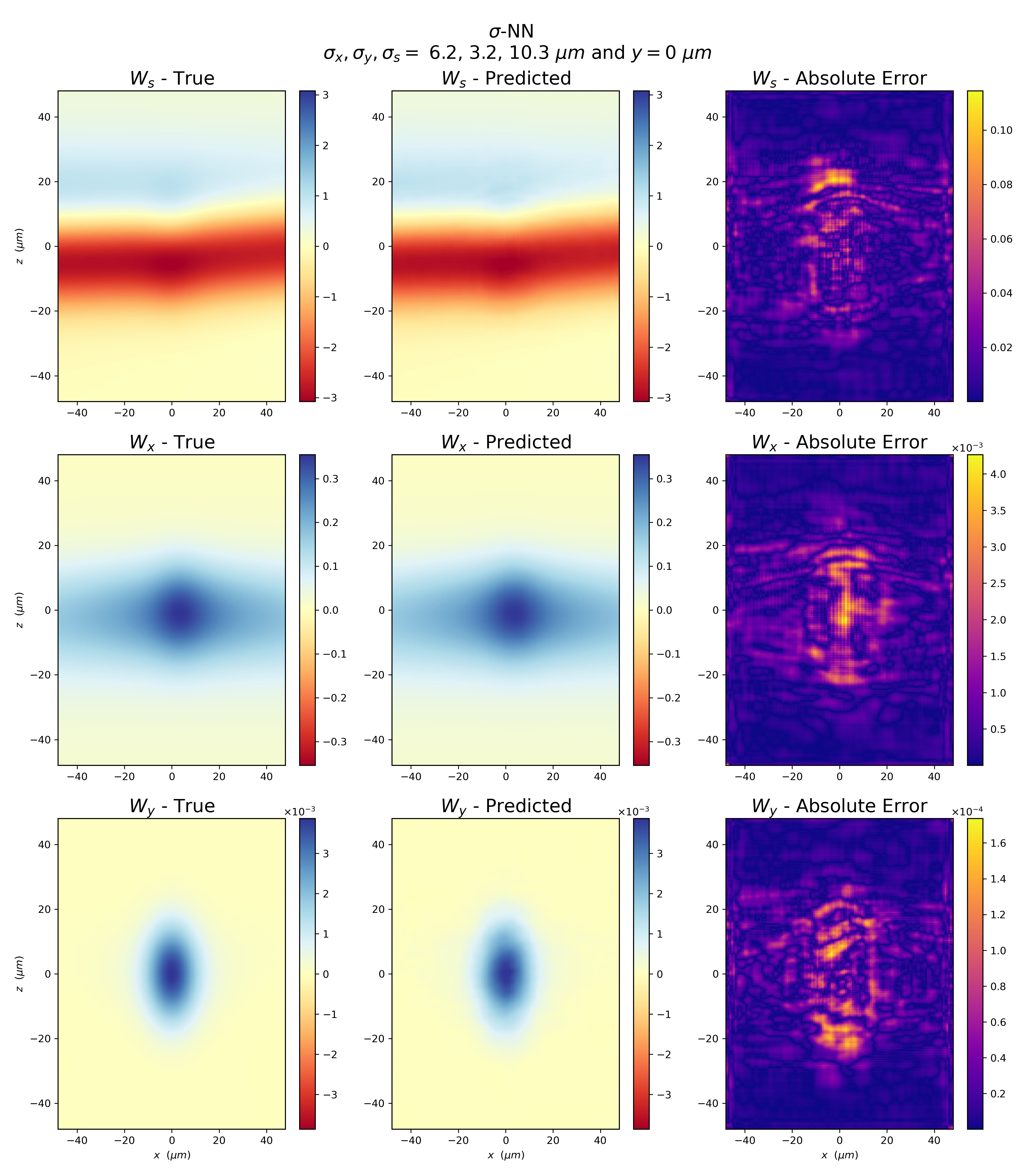}
\caption{\textbf{Gaussian Dataset}: $\sigma$-NN  performance on a distribution in the Test Set in the $y = 0 ~ \mu m$ plane. Wake units: $10^6 /m^2$.
}
 \label{Sigma-NN-2D}
\end{figure*}

\begin{figure*}
\centering
\includegraphics[ width=0.99\textwidth ]{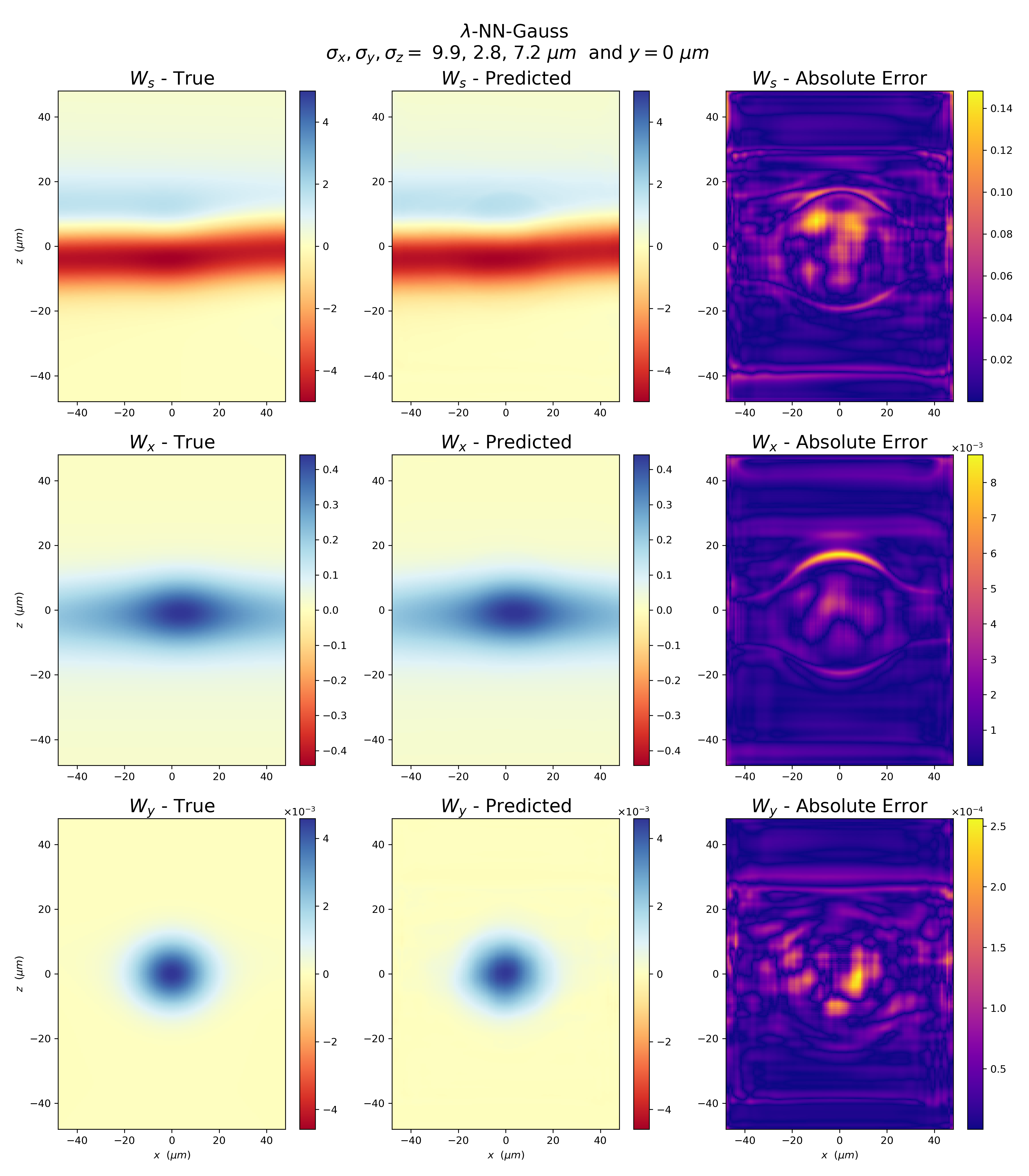}
\caption{\textbf{Gaussian Dataset}: $\lambda$-NN-Gauss performance on a distribution in the Test Set in the $y = 0 ~ \mu m$ plane. Wake units: $10^6 /m^2$. 
}
 \label{Lambda-NN-2D}
\end{figure*}

\begin{figure*}
\centering
\includegraphics[ width=0.99\textwidth ]{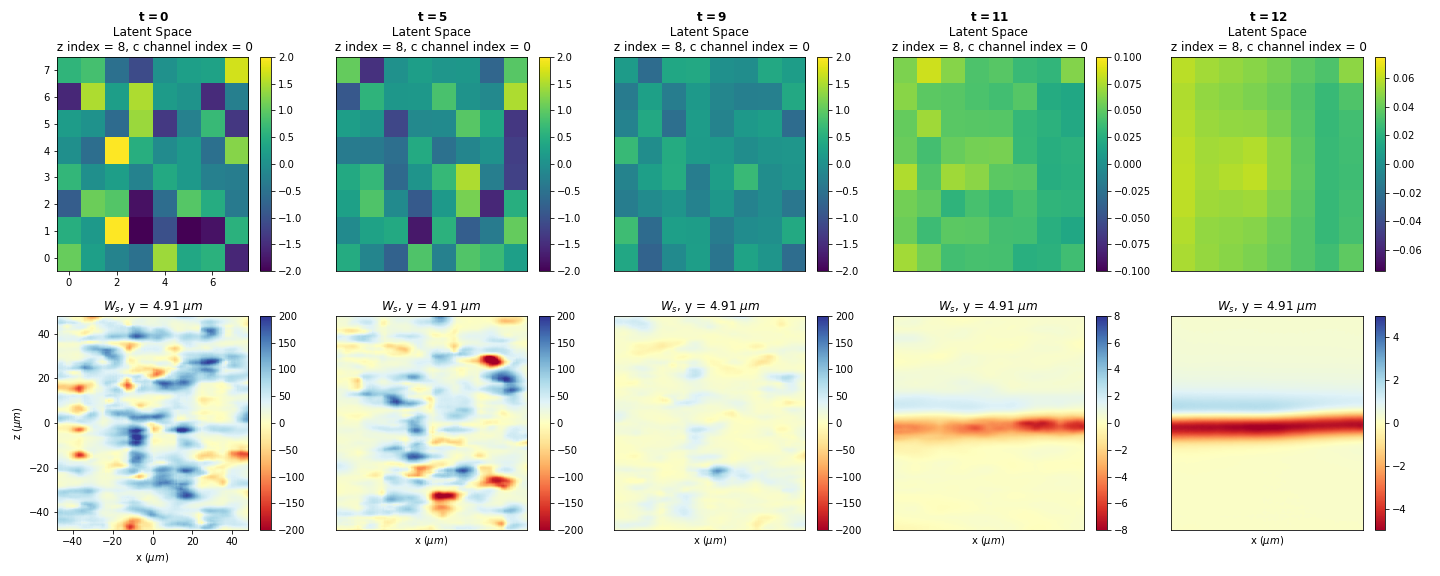}
\caption{\textbf{Gaussian Dataset}: Latent conditional diffusion. \textit{Top}: A cross-section of latent space during the denoising process at certain time steps. \textit{Bottom}: A cross-section of the corresponding wakefield during denoising. Wake units: $10^6 /m^2$. }
 \label{LCD-Evolution}
\end{figure*}

To evaluate the models, we used the relative weighted mean average error (RW-MAE) for each component $i=x, y, s$:
\begin{align}
\text{RW-MAE}_i = \frac{\int d^3\mathbf{r} \lambda(\mathbf{r})|W^i_{pred.}(\mathbf{r})-W^i_{true}(\mathbf{r})|}{\int d^3\mathbf{r} \lambda(\mathbf{r})|W^i_{true}(\mathbf{r})|},
\end{align}
where $W^i_{true}(\mathbf{r})$ and $W^i_{pred.}(\mathbf{r})$ are the true and predicted wakefield, respectively. The idea behind RW-MAE is that we care most about being correct where the particle density is highest, since $\mathbf{W}(\mathbf{r})$ is being used to update the phase space distribution. Moreover, it  differs significantly enough from the MSE loss function used in training to serve as a complementary metric.

\subsection{Gaussian Dataset}

\begin{table*}[]
\centering
\begin{tabular}{llccc}
\hline
\textbf{Model}                      & \textbf{Set}                                                            & \begin{tabular}[c]{@{}c@{}}$\textbf{s}$\\ Mean (STD), \%\end{tabular}      & \begin{tabular}[c]{@{}c@{}}$\textbf{x}$ \\ Mean (STD), \%  \end{tabular} & \begin{tabular}[c]{@{}c@{}}$\textbf{y}$\\ Mean (STD), \%  \end{tabular} \\ \hline
\multirow{2}{*}{$\sigma$-NN}        & Training \quad \quad \quad \quad & 1.24 (0.54) \quad \quad \quad \quad & 0.59 (0.10) \quad \quad \quad \quad & 1.09 (0.18)                                                                \\
                                    & Test \quad \quad \quad \quad     & 1.26 (0.65) \quad \quad \quad \quad & 0.60 (0.13) \quad \quad \quad \quad & 1.12 (0.17)                                                                \\ \hline
\multirow{2}{*}{$\lambda$-NN-Gauss} & Training \quad \quad  \quad \quad  & 1.39 (1.18) \quad \quad  \quad \quad  & 0.48 (0.33) \quad \quad  \quad \quad  & 0.97 (0.62)                                                                \\
                                    & Test \quad \quad  \quad \quad      & 1.43 (1.33) \quad \quad  \quad \quad  & 0.49 (0.39) \quad \quad  \quad \quad  & 0.97 (0.61)                                                                \\ \hline
\multirow{2}{*}{LCD} & Training \quad \quad  \quad \quad  & 6.59 (4.32) \quad \quad  \quad \quad  & 2.78 (1.82) \quad \quad  \quad \quad  & 5.35 (3.27)                                                                \\
                                    & Test \quad \quad  \quad \quad      & 6.68 (4.22) \quad \quad  \quad \quad  & 2.93 (1.92) \quad \quad  \quad \quad  & 5.43 (2.45)                                                                \\ \hline
\end{tabular}
\caption{\textbf{Gaussian Dataset}: The $\text{RW-MAE}_i$ over the training and test set. }\label{Gaussian-results}
\end{table*}



We compare the results of the ML models to the approximate 1D theory \cite{Saldin:1996gs}, which can be obtained by using a Gaussian distribution in Eq. (\ref{1D-theory-Ws}):
\begin{align}
W_s(z) &= \frac{2^{11/6}}{3^{7/6}\rho^{2/3}\sigma_s^{4/3}} \Bigg[\frac{\sqrt{\pi} z {}_{1} F_1  \left(\frac{7}{6}, \frac{3}{2}, -\frac{z^2}{2}) \right)}{\sqrt{3} \Gamma(\frac{5}{3})} 
 \\
&- \frac{2^{5/6}  \Gamma(\frac{2}{3}) {}_{1} F_1  \left(\frac{2}{3}, \frac{1}{2}, -\frac{z^2}{2}) \right)}{\Gamma(\frac{7}{3})} \Bigg], \nonumber
\end{align}\label{Ws-1D-theory}
where ${}_{1} F_1  \left(a, b, z \right)$ is a confluence hypergeometric function of the first kind. While the networks don't produce a perfect match, both were generally closer to the $W_s$ produced by \texttt{PyCSR3D} than the 1D theory frequently used in practice, as can be seen in Fig. \ref{CSR- lambda - NN - 1D - General}.

For the horizontal wake in the 1D approximation, various authors have arrived at $W_x(z) = \frac{\Lambda}{\rho }  \lambda(z)$, with different sources arriving at different values for $\Lambda$, depending on the assumptions used. For a Gaussian distribution :

\begin{align}\label{Wx-1D-theory}
W_x(z) =  \frac{\Lambda}{\rho }  \frac{1}{\sqrt{2\pi \sigma_s^2 }} e^{-z^2/2\sigma_s^2 }, 
\end{align}
While some authors arrive at constant values, such as $\Lambda = 2$ and $\Lambda = 4$ \cite{Derbenev:1995eg, Cai:2020kvm},  we found that:
\begin{align}\label{Lambda-used}
\Lambda = \log \left[ \left( \frac{\rho \sigma_s^2}{\sigma_\perp^3}\right)^{2/3} \left(1 + \frac{\sigma_\perp}{\sigma_s} \right) \right],
\end{align}
from Ref. \cite{Derbenev:1996ub} with $\sigma_\perp= \sqrt{\sigma_x^2 + \sigma_y^2}$ worked the best at approximating the \texttt{PyCSR3D} wakefields. Again, the predictions of $\sigma$-NN and $\lambda$-NN were both generally closer to \texttt{PyCSR3D} than the 1D theory, an example of which can be seen in Fig. \ref{Sigma-Lambda-NN-1D}. It was observed that the tails generally drop much slower than Eq. (\ref{Wx-1D-theory}) suggests. 

\begin{table}[]
\centering
\begin{tabular}{ll}
\hline
\textbf{Model}                & \begin{tabular}[c]{@{}l@{}} \textbf{Evaluation Time} \\  Mean (STD), seconds\end{tabular} \\ \hline
\texttt{PyCSR3D} & 99.2 (2.2)                                                                                                  \\
LCD                           & 2.94 (0.09)                                                                                                   \\
$\lambda$-NN-Gauss            & 0.410 (0.051)                                                                                               \\
$\sigma$-NN                   & 0.106 (0.002)                                                                                        
\end{tabular}
\caption{\textbf{Gaussian Dataset}: Time to predict $\mathbf{W}(\mathbf{r})$ for a single distribution, from highest to lowest. The mean and standard deviation are derived from 7 trials.}\label{Time-Gaussian}
\end{table}

As can be seen in Table \ref{Gaussian-results}, the $\sigma$-NN model and $\lambda$-NN-Gauss performed comparably across the $\text{RW-MAE}_i$ metric. Thus, despite $\sigma$-NN having an easier task of just predicting the wakefields from the given $\sigma_i$'s, its accuracy and that of $\lambda$-NN were similar. The addition of attention layers and residual connection were invaluable, as testing showed that a simpler CNN architecture without these features increased the error of $\lambda$-NN by a factor $\sim$3. However, the performance of $\lambda$-NN-Gauss comes at the cost of taking $\sim$4 times longer to evaluate than $\sigma$-NN (see Table \ref{Time-Gaussian}), likely due to the greater model complexity.

The LCD performed worse than the two other ML models, both in terms of accuracy and evaluation time. The accuracy result is likely due to having more sources of potential error: encoding the distribution, running the backwards diffusion and then decoding to get $W^i(\mathbf{r})$. Similarly, the use of several networks and having to perform backward diffusion makes the process slower than the other models. 

Nonetheless, all the machine learning models achieved their predictions much faster than \texttt{PyCSR3D}. For the time to predict $\boldsymbol{W}$ from a single distribution, $\sigma$-NN saw a speedup of $\sim$1000, $\lambda$-NN-Gauss  $\sim$250 and LCD $\sim$34 (see Table \ref{Time-Gaussian}). 
The most computational time-consuming part for \texttt{PyCSR3D} was calculating grid values for the Green's functions.


\subsection{Multi-Gaussian Dataset}

\begin{table*}[]
\centering
\begin{tabular}{llccc}
\hline
\textbf{Model}                & \textbf{Set} &  \begin{tabular}[c]{@{}c@{}}$\textbf{s}$\\ Mean (STD), \%\end{tabular}      & \begin{tabular}[c]{@{}c@{}}$\textbf{x}$ \\ Mean (STD), \%  \end{tabular} & \begin{tabular}[c]{@{}c@{}}$\textbf{y}$\\ Mean (STD), \%  \end{tabular} \\ \hline

\multirow{2}{*}{$\lambda$-NN-Multi} & Training  \quad \quad \quad     & 3.60 (1.51)  \quad \quad \quad      & 0.96 (0.41)   \quad \quad \quad     & 2.56 (0.76)         \\
                              & Test  \quad \quad \quad          & 3.60 (1.51)  \quad \quad \quad      & 0.99 (0.56)  \quad \quad \quad      & 2.74 (0.92)        
\end{tabular}
\caption{\textbf{Multi-Gaussian Dataset}: The average of $\text{RW-MAE}_i$ over the training/test set.}\label{General-results}
\end{table*}

\begin{figure}
\centering
\includegraphics[ width=0.98\columnwidth ]{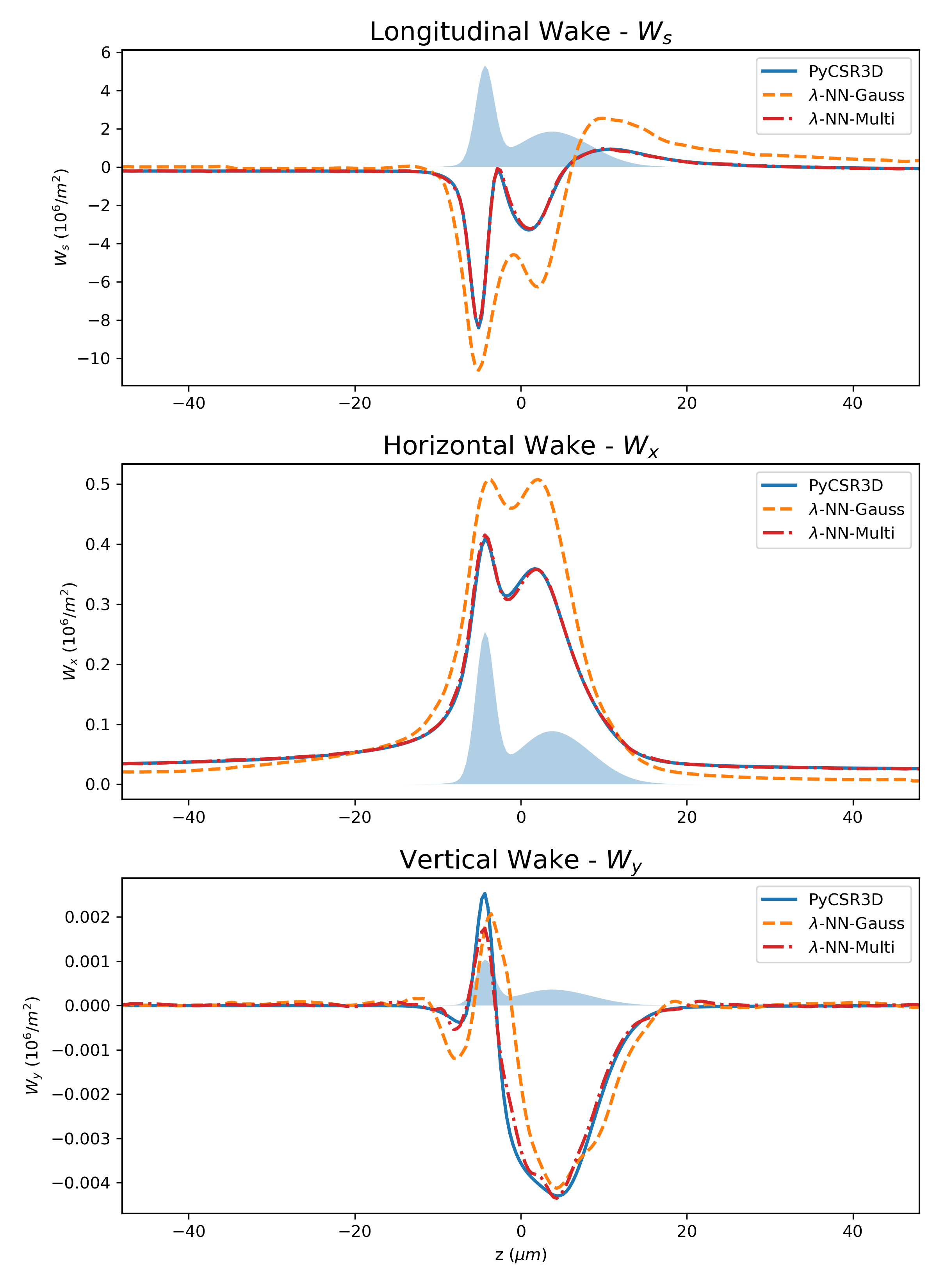}
\caption{\textbf{Multi-Gaussian Dataset}: The performance of $\lambda$-NN-Gauss and  $\lambda$-NN-Multi on a distribution in the Test Set. This is along the z axis ($x = y = 0 ~ \mu m$), with the electron distribution along the axis shown in the shaded blue area. 
}
 \label{CSR- lambda - NN - 1D - General}
\end{figure}

\begin{figure*}
\centering
\includegraphics[ width=0.95\textwidth ]{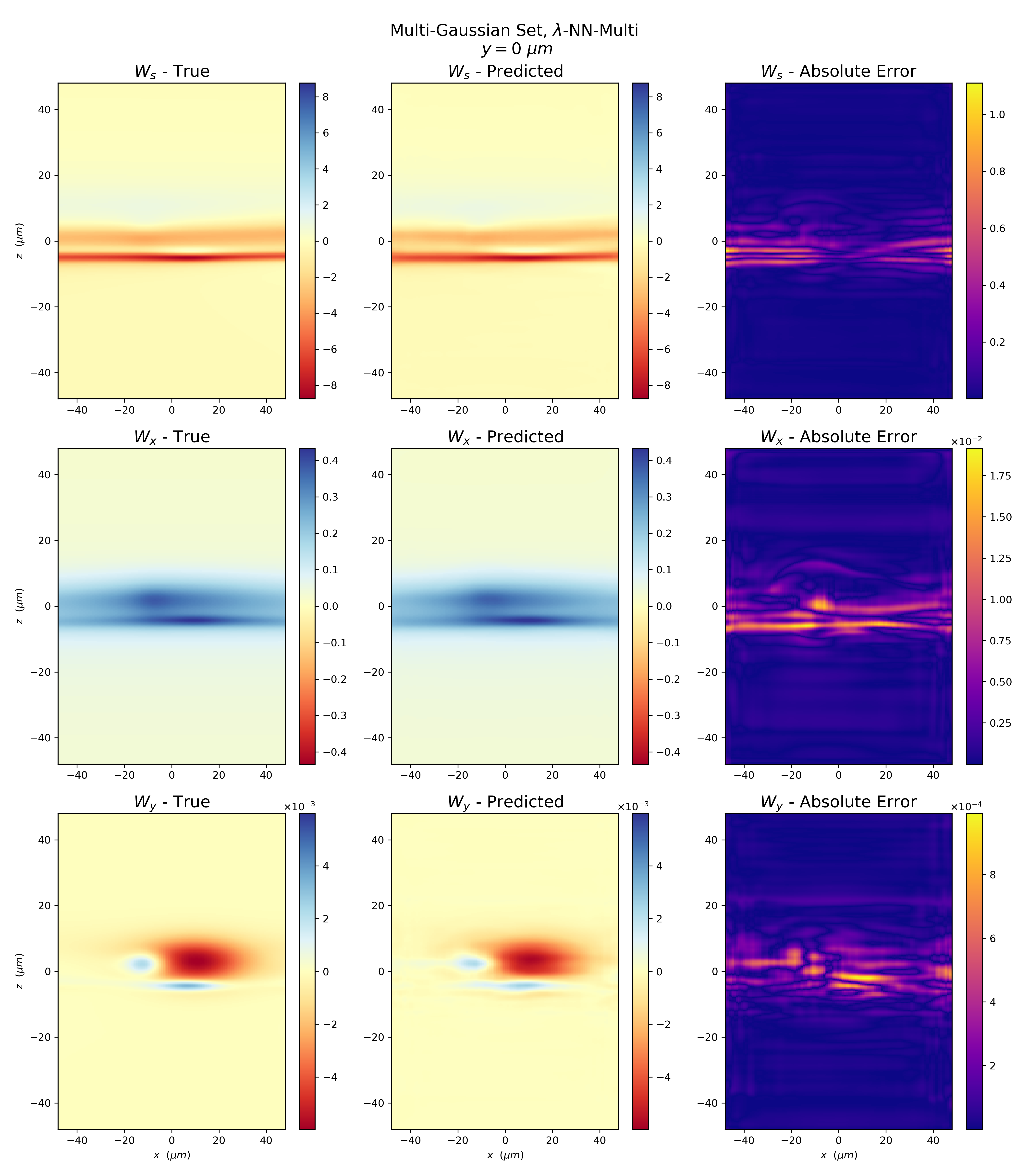}
\caption{\textbf{Multi-Gaussian Dataset}: $\lambda$-NN-Multi performance on a distribution in the Test Set on the $y = 0 ~ \mu m$ plane.  Wake units: $10^6 /m^2$. }
 \label{CSR- lambda - NN - 2D - General}
\end{figure*}

The $\lambda$-NN-Gauss model was able to qualitatively predict the wakes in the Multi-Gaussian Dataset, as can be seen in Fig. \ref{CSR- lambda - NN - 1D - General}, despite only ever being trained on single Gaussian distributions. 
Given this result, researchers using CNN models for CSR calculations can greatly benefit from pre-training their models on easily producible, large datasets using analytic results from approximate models, before fine-tuning them on more realistic and general data from numerically intensive simulations. Since much learning transferred, training for $\lambda$-NN-Multi took half the time than training $\lambda$-NN-Gauss from scratch. Nonetheless, $\lambda$-NN-Multi benefited greatly from training on Multi-Gaussian Dataset.

Evaluation times for the $\lambda$-NN-Multi and \texttt{PyCSR3D} were similar to what was found with the Gaussian Dataset in Table \ref{Time-Gaussian}.

\subsection{Extrapolation and Out of Distribution Generalization}

\begin{figure}
\centering
\includegraphics[ width=0.97\columnwidth ]{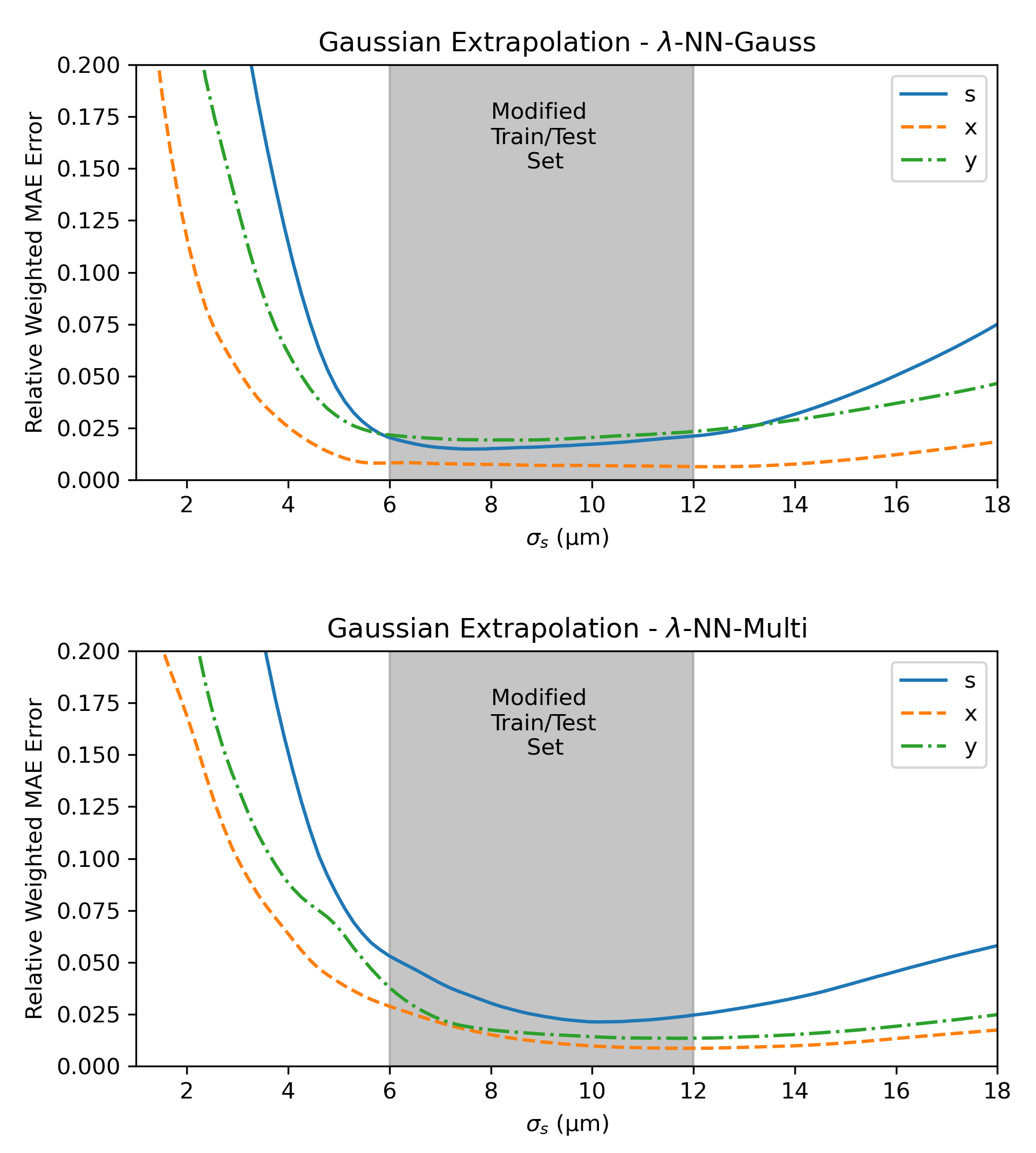}
\caption{Extrapolation to Gaussians with different $\sigma_s$ from the Gaussian training/test data.}
 \label{OOD Sigma z}
\end{figure}

\begin{figure}
\centering
\includegraphics[ width=0.97\columnwidth ]{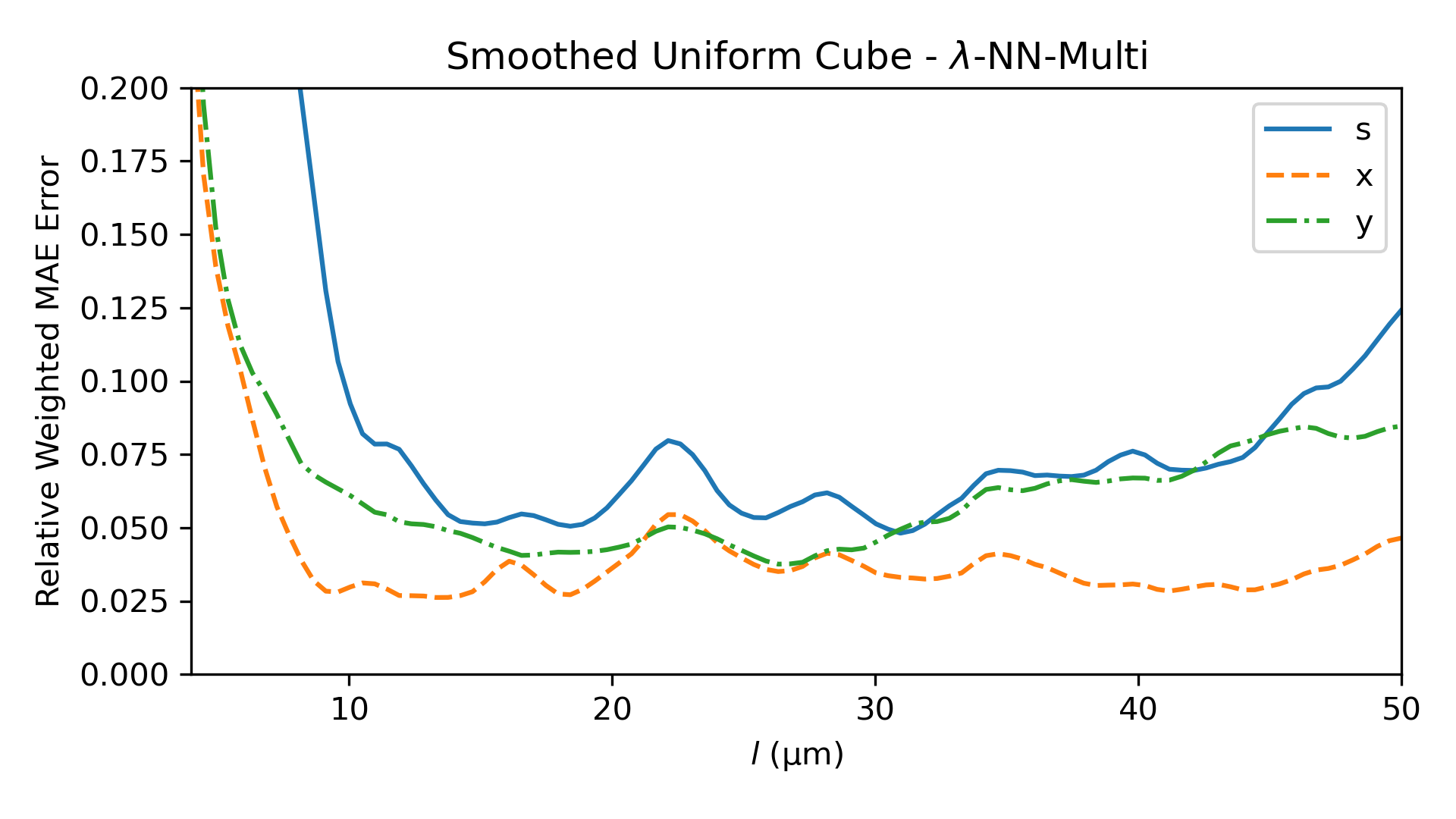}
\caption{Smooth Uniform Cube - RW-MAE error for each component as a function of cubic side length, $l$.}
 \label{OOD UNIFORM Cube}
\end{figure}

To see how well $\lambda$-NN generalizes to  distributions very different from those in the training/test data, we performed two different tests.  
\subsubsection{Gaussian Extrapolation}

First, we evaluated how well $\lambda$-NN-Gauss and $\lambda$-NN-Multi extrapolate by using Gaussian distributions with widths smaller/larger than what the model had encountered before. Since our lattice spacing for the transverse (longitudinal) component is $0.75 ~\mu m$ ($0.375 ~\mu m$) and the lowest width in the training data is $\sigma_i = 2 ~\mu m$, it would be difficult to tell if failure to extrapolate to smaller $\sigma_i$ stems from the model being unable to extrapolate or the small number of samples due to the lattice spacing. We thus modified the training/test data to the range $ 6 \leq \sigma_i \leq 12 ~\mu m$ ($i= x, y, s$) for both the Gaussian and Multi-Gaussian Dataset, keeping the number of samples in the sets the same. A new $\lambda$-NN's were then trained from scratch on the modified datasets, using the same procedure as before. 

Next, a new dataset was created, which consisted of Gaussians with spreads that went outside the training/test set. Specifically, the longitudinal spread went from $\sigma_s = 1$ to $18 ~\mu m$, while the transverse spreads were kept fixed at $\sigma_x = \sigma_y = 9 ~\mu m$. 

As can be seen in Fig. \ref{OOD Sigma z}, the $\lambda$-NN-Multi model performed well at extrapolating to larger $\sigma_s$ with the RW-MAE$_i$ error for each component only rising slowly outside training/test data. In contrast, the predictions for smaller $\sigma_s$ did much worse and rose rapidly as $\sigma_s$ was decreased. This occurred far before $\sigma_s$ approached the $z$ lattice length of $0.375  ~\mu m$, thus is likely due to the model's failure to extrapolate to lower $\sigma_s$.

\subsubsection{Generalization: Smoothed Uniform Cubical Distribution}

A 1D uniform distribution between two points, $a$ and $b$, can be considered the difference of two step functions, each  centered at $a$ and $b$. This can be smoothed by replacing the step function with sigmoids: $f(x) = \frac{1}{b-a}\left(\frac{1}{1+e^{x-a}}-\frac{1}{1+e^{x-b}}\right)$. We call a smoothed uniform cubical distribution one that uses $f(x)$ for all 3 components with the same widths (note: it's not exactly a cube in physical space due to curvilinear coordinates being used, but due to its small size relative to $\rho$ it is only approximately cubical). Smoothing is done to improve the numerics in the simulation, which depend on derivatives for the longitudinal component. Also, a smooth drop-off in density is a more realistic description of particle beams than a discontinuous drop. While the sampling done in the Multi-Gaussian is from a cubical distribution, the relatively low number of samples, 25, the smoothed edges and the fact that the cube length varied means the Smooth Uniform Cubical Distributions are quite different from what the model has seen.

The $\lambda$-NN-Multi model from the previous subsection was used. Going from side lengths of $l=4 ~\mu m$ to $l = 50 ~ \mu m$, the RW-MAE can be seen in Fig. (\ref{OOD UNIFORM Cube}). The results are similar to the Gaussian case, where the models fail at small widths, but become almost level at larger ones. Based on this and the previous results, it seems the $\lambda$-NN-Multi should only be used on distributions with widths larger than $\sim 10 ~\mu m$.

\section{Conclusion}
We have created surrogate ML models to find the 3D CSR wakes. For the  $\lambda$-NN and $\sigma$-NN models, for the price of a percent level error, we get an increase in speed ranging from $\sim$250 and $\sim$ 1000, respectively.

Now that it has been shown these ML models have the ability to produce the wakefields accurately and quickly, it would be interesting to apply to real accelerators. One interesting case would be  emittance exchangers, where CSR plays a large role.

A long term goal would be to make a computationally efficient, self-consistent general CSR code for the full 6D phase space for arbitrary motion. 

A number of modifications can be made to make the models to further improve performance and applicability of the models. One issue is that by using a CNN architecture, we have introduced a mesh-dependence. Finding ways to embed the data, such as is done in Fourier Neural Operators \cite{li2021fourier}, can get rid of the mesh dependence. We have neglected shielding effects from the surrounding metallic beam pipe. In some cases these may be important and future work can look into including these. Transience has also been neglected here. The 2D theory already exists \cite{Cai:2021xxm} and ML has been applied to that case \cite{Robles:2023tus}, so a fully 3D transient case would be the next goal. 


\bibliographystyle{apsrev}

\end{document}